\documentstyle[12pt,epsf]{article}
\newcommand{\beq}{\begin{equation}}
\newcommand{\eeq}{\end{equation}}
\newcommand{\bea}{\begin{eqnarray}}
\newcommand{\eea}{\end{eqnarray}}

\newcommand{\CZ}{{\cal Z}}

\newcommand{\gstr}{\gamma_{str}}
\newcommand{\pa}{\partial}

\newcommand{\nn}{\nonumber}

\begin{document}
\topmargin 0pt
\oddsidemargin 5mm
\headheight 0pt
\topskip 0mm

\addtolength{\baselineskip}{0.20\baselineskip}

\pagestyle{empty}

\begin{flushright}
OUTP-97-30P\\
25th August 1997\\
hep-th/9708027
\end{flushright}

\begin{center}

\vspace{18pt}
{\Large \bf Branched Polymers, Complex Spins and the Freezing Transition}

\vspace{2 truecm}

{\sc Jo\~ao D. Correia\footnote{e-mail: j.correia1@physics.ox.ac.uk},
Behrouz Mirza\footnote{e-mail: b.mirza@cc.iut.ac.ir}
and John F. Wheater\footnote{e-mail: j.wheater1@physics.ox.ac.uk}}

\vspace{1 truecm}

{\em Department of Physics, University of Oxford \\
Theoretical Physics,\\
1 Keble Road,\\
 Oxford OX1 3NP, UK\\}

\vspace{3 truecm}

\end{center}

\noindent 
{\bf Abstract.} We show that by coupling complex three-state systems to 
branched-polymer like ensembles we can obtain models with $\gstr$
different from $\frac{1}{2}$. It is also possible to study the
interpolation between dynamical and crystalline graphs for these models;
we find that only when geometry fluctuations are completely forbidden is
there a crystalline phase. These models share many of the properties of
full two-dimensional quantum gravity.

\vfill
\begin{flushleft}
PACS: 04.60.Nc,5.20.-y,5.60.+w\\
Keywords: conformal matter, quantum gravity, branched polymer\\
\end{flushleft}
\newpage
\setcounter{page}{1}
\pagestyle{plain}

\pagestyle{plain}

\section{Introduction}

Recent years have seen remarkable progress in the theory of two
dimensional quantum gravity, due in large extent to the discretized
approach (see \cite{ginsparg}, \cite{revambj} for a review). We have a
good understanding of the situation for $c < 1$, where $c$ is the
central charge of the matter coupled to gravity, through the KPZ formula
\cite{kpz} and associated results. The situation changes for $c>1$
where the KPZ formula gives meaningless predictions; numerical
simulations  \cite{num} and analytical arguments \cite{harris},
\cite{me}, \cite{david} favour the view that quasi one-dimensional 
configurations
called branched  polymers dominate the statistical ensemble in this
region (there is, however, no definite proof of this conjecture). For
generic branched polymer ensembles the critical exponent
$\gstr=\frac{1}{2}$.
Certain modifications of the weighting of different branched polymers lead to 
other positive values of $\gstr$ (\cite{polyising}, \cite{pialas}). 

The KPZ results show that the critical behaviour of spin systems (e.g.
Ising spins) interacting with two-dimensional quantum gravity is
modified from that observed on a regular lattice. This raises the
question of what would happen in an extended theory which interpolates
between the fixed lattice (or crystalline) model and the gravity model
\cite{jfwferg}. The only two-dimensional problem of this type that has
been solved is the case of $R^2$ gravity (here $R$ is the scalar
curvature density) \cite{r2}. If surfaces of large $R^2$ are suppressed
the ensemble becomes flatter and flatter at short distance scales.
 However it is found that
$\gstr=-\frac{1}{2}$, the pure gravity value, for all finite values of
the $R^2$ coupling; only when all surfaces with non-minimal $R^2$ are
completely forbidden does $\gstr$ change. 

In this paper we will show that it is possible to change $\gstr$ for a
branched polymer (BP) ensemble by introducing matter fields, rather than
fiddling directly with the geometry, just as matter fields at
criticality can change $\gstr$ in the two-dimensional models. Then we
study the interpolation between ``gravity'' and ``crystalline'' phases.

As a first step, in
Section 2 we classify the possible critical exponents obtainable from a
generic class of quadratically coupled non-linear equations. In Section 3
we introduce and study the set of binary trees. Section 4 is concerned with the 
coupling of matter to binary trees. 
First we show that a normal Ising model coupled  to binary trees yields
the same results as pure trees.
 Introducing a more complicated model of ${\bf Z}_3$ spins with a
complex action leads to new behaviour, with $\gstr$ taking values of
$2$, $\frac{3}{2}$ and $\frac{2}{3}$ in addition to $\frac{1}{2}$.

\section{Categorization of Critical Behaviour}

In what follows, we will be interested in the possible critical behaviour
of systems of coupled equations of the form
\beq A_i = z + \alpha_{i}^{jk} A_j A_k. \label{ai} \eeq
The $A_i$ are grand canonical ensemble partition functions and $z$ is
related to the chemical potential $\mu$ by $z=e^{-\mu}$.
The coefficients $\alpha_{i}^{jk}$ depend on the coupling constants of
the specific model and the repeated indices are summed over; we assume that the set of equations is irreducible in the sense that no linear combination
of $A_i$s decouples. At small
$z$ the $A_i$ are power series in $z$; as $z$ increases the $A_i$ become
non-analytic at some critical value $z_c$. Defining
\beq M_{ij} = \delta_{ij} - \alpha_i^{jk} A_k - \alpha_i^{jk} A_j
\label{defM} \eeq
and differentiating equation (\ref{gen}) w.r.t. $z$
we obtain
\beq \frac{\pa A_i}{\pa z} = \left ( M^{-1} \right )_{ij} u^i
\label{diff} \eeq
where $u^j=(1$,...,$1)$ and  $M^{-1}$ is the inverse of $M$, $M^{-1} =
\frac{1}{det(M)} (cof(M))^{\bf T}$. There are two ways in which the above
expression can become non-analytic; either the $A_i$ remain finite as $z
\uparrow z_c$ but their derivatives blow up, or the $A_i$ themselves
blow up in that limit. We  analyse of these two cases separately.

If the $A_i$ remain finite,  but the $\frac{\pa A_i}{\pa z}$ diverge, then
we may write, with $\zeta = z_c - z$,
\beq A_i (z) = R_i(\zeta) - \zeta^{1-\gstr} P_i(\zeta) \label{finai} \eeq
where $\gstr <1$ and  $R_i(\zeta)$ and  $P_i(\zeta)$ are regular functions
 at $\zeta=0$ with  $P_i(0) \neq 0$, $R_i(0) \neq 0$. Then 
\beq \frac{\pa A_i}{\pa z} \sim \zeta^{- \gstr} \label{dfinai} \eeq
Now, since at $z=z_c$ the $A_i$ remain finite, we must have $det(M)=0$ for
the derivatives to diverge. The general expression for the determinant is
obtained by substituting
 (\ref{finai}) into (\ref{defM}) and expanding which gives
\beq det M = \zeta^{1-\gstr} P^{(1)} (\zeta) + \zeta^{2(1-\gstr)} P^{(2)} (\zeta)
+ \ldots + \zeta^{n(1-\gstr)} P^{(n)} (\zeta). \label{expdetM} \eeq
The $P^{(n)} (\zeta)$ depend on the $\alpha_i^{jk}$ and therefore on the
coupling constants.

We can now identify the different cases, starting with 
the most general; all the $P^{(n)} (0)$ are nonzero. By comparing the leading powers
in (\ref{expdetM}) and (\ref{dfinai}) we see that $\gstr$ obeys 
\beq -\gstr = -(1-\gstr) \eeq
and therefore $\gstr=\frac{1}{2}$. No constraints on the coupling
constants are necessary for this case; so for example this value of $\gstr$
can hold for areas of the phase diagram of a system with two coupling constants.

The next case is that of $P^{(1)}(0)=0$ but $P^{(2)}(0) \neq 0$. Then the expansion
of the determinant (\ref{expdetM}) becomes
\beq det M \sim  \zeta^{2(1-\gstr)}+O(\zeta^{2-\gstr})  + O(\zeta^{3(1-\gstr)})
+ \mbox{...} \label{pzero} \eeq
and the leading singularity of (\ref{pzero})  is in the
first term, leading to
\beq -\gstr = -2 (1- \gstr) \eeq
and to a value of $\gstr=\frac{2}{3}$. Notice that in order to make
$P^{(1)}(0)=0$ we are introducing a constraint in the coupling space. In the
case of two coupling constants  this
leads to a line in the phase diagram.

Further values of $\gstr$ will be obtained if we introduce more
constraints.  If we demand $P^{(1)} (0) = P^{(2)} (0) =0$ but allow $P^{(3)} (0) \neq
0$, the leading term will be the $O(\zeta^{3(1-\gstr)})$ and we obtain
$\gstr=\frac{3}{4}$; the two constraints necessary to achieve this
situation mean that, in the case of two coupling constants, this value of
$\gstr$ is confined to a point in the phase diagram.
A higher number of constraints will lead to other values of $\gstr$, but
the above instances are enough for our purposes. Note that not all of the possible exponents found in this way can arise from any set of equations; the
larger the number of coupled equations the more exotic is the possible behaviour.

The second case is that the $A_i$ themselves become infinite as
$z \uparrow z_c$; of course the derivatives will also blow up. In this
case
\beq A_i (z) = \zeta^{1-\gstr} P_i(\zeta) + R_i(\zeta)
\label{infinai} \eeq
with $\gstr>1$. Again  $det M$ must vanish at $z = z_c$. Putting (\ref{infinai}) into
(\ref{expdetM}) and expanding yields
\beq det M \sim \sum_i \zeta^{(1-\gstr) m_i} \zeta^{n_i} a_i \label{mn}
\eeq
where $m_i$,$n_i$ are positive integers and $a_i$ are (coupling constant 
dependent) constants such that 
\beq m_i (1-\gstr) + n_i > 0 \label{constmn} \eeq
so that we get 
\beq - \gstr = -(m_j (1- \gstr) + n_j) \label{cond} \eeq
if $a_1 = a_2 = \mbox{...} = a_{j-1} = 0$ and $a_j \neq 0$.
By assumption $\gstr>1$ so that we must always have $n_i > 0$. If $n_i=1$ then 
$\gstr=1$ (otherwise from (\ref{constmn}) $m_i$ is negative, which is
forbidden); this implies logarithmic singularities in $A_i$ which are
not possible from a set of polynomial equations so we conclude that $n_i
>1$.
The
question is now how many $a_i$ must be zero (i.e., how many constraints
there are) to get a certain value of $\gstr$.

The number of constraints necessary to get different  values of $\gstr$
can be obtained by considering equation (\ref{constmn}). For instance, if
we demand $\gstr=\frac{3}{2}$, (\ref{constmn}) reduces to $n_i -
\frac{m_i}{2} > 0$. We can then examine the possibilites for $n$ and $m$.
Setting $m=1$, we can put $n=1$ (giving a value of $\frac{1}{2}$) and $n=2$
(giving $\frac{3}{2}$); we can also put $n=m=2$ giving a value of 1. Thus
to obtain $\gstr=\frac{3}{2}$ we need a minimum of two constraints.
The results of this analysis for some values of $\gstr$ are shown in Table 1.

\begin{table}
\caption{Constraints and $\gstr$}
\begin{center}
\begin{tabular}{|c|c|c|c|c|}\hline
$\gstr$ & Constraint & Possible Values & $(n$,$m)$ & $\#$ of constraints
\\ \hline \hline
$\frac{3}{2}$ & $n_i - \frac{m_i}{2} > 0$ & $\frac{1}{2}$ & $(1$,$1)$ & \\
 & & 1 & $(2$,$2)$  & 2 \\ 
 & & $\frac{3}{2}$ & $(2$,$1)$ & \\ \hline  
$\frac{5}{4}$ & $n_i-\frac{m_i}{4} > 0$ & $\frac{3}{4}$ & $(1$,$1)$ &  \\
 & & $\frac{1}{2}$ & $(1$,$2)$ & \\ 
 & & $\frac{1}{4}$ & $(1$,$3)$ & 4 \\ 
 & & 1 & $(2$,$4)$ & \\ 
 & & $\frac{5}{4}$ & $(2$,$3)$ & \\ \hline
2 & $n_i - m_i > 0$ & 1 & $(2$,$1)$ & 1 \\ 
 & & 2 & $(3$,$1)$ & \\ \hline
3 & $n_i - 2 m_i > 0$ & 1 & $(3$,$1)$ & \\ 
 & & 2 & $(4$,$1)$ & 2\\ 
 & & 3 & $(5$,$1)$ & \\ \hline
4 & $n_i - 3 m_i > 0$ & 1 & $(4$,$1)$ & \\ 
 & & 2 & $(5$,$1)$ & \\ 
 & & 3 & $(6$,$1)$ & 3 \\ 
 & & 4 & $(7$,$1)$ & \\ \hline
\end{tabular}
\end{center}
\end{table}

\section{Binary Trees}

The properties of the ensemble of tree graphs are well known \cite{kaz}.
In this paper we will consider trees made of cubic vertices but modify the 
graphs slightly so that all the external lines except the root are
attached to another line.
\begin{figure}
\begin{center}\parbox{10cm}{
\epsfxsize 10cm
\epsfbox{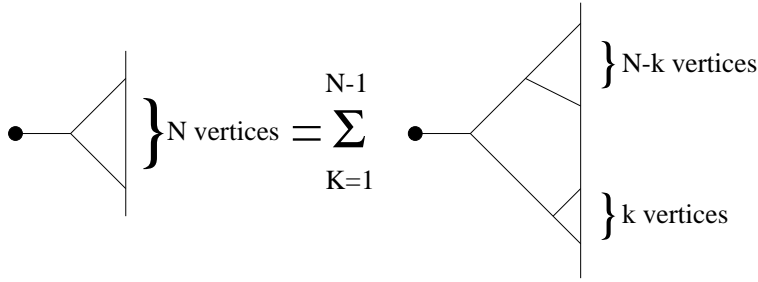}}\end{center}
\caption{The recurrence relation for tree graphs.}
\label{iter}
\end{figure}
This does not affect the recurrence relation  generating the graphs
which is shown in fig.\ref{iter}. Letting $T_N$ be the number of graphs with $N$
external vertices (not counting the root) we have
\beq T_N = \sum_{k=1}^{N-1} T_{N-k} T_k \mbox{\ , $T_1=1$}
\label{recurr} \eeq
or
\beq T(z) = \sum_{N=1}^{\infty} z^N T_N = \frac{1}{2} \left ( 1-
\sqrt{1-4 z} \right ). \label{gen} \eeq
The exponent $\gstr$ for the ensemble of graphs with one marked point
(the root in this case) is defined so that the generating function $G$
for the number of graphs of a given size has leading non-analytic
behaviour
\beq \frac{\pa G}{\pa z} = (z_{cr}-z)^{-\gstr} \label{leading} \eeq
so for the tree ensemble $\gstr=\frac{1}{2}$. Now we want to modify the
problem so that one particular sort of graph is picked out and given a
different weight; this is the ladder graph shown in fig.\ref{ladder}.
\begin{figure}[h]
\parbox{7cm}{
\epsfxsize 5cm
\epsfbox{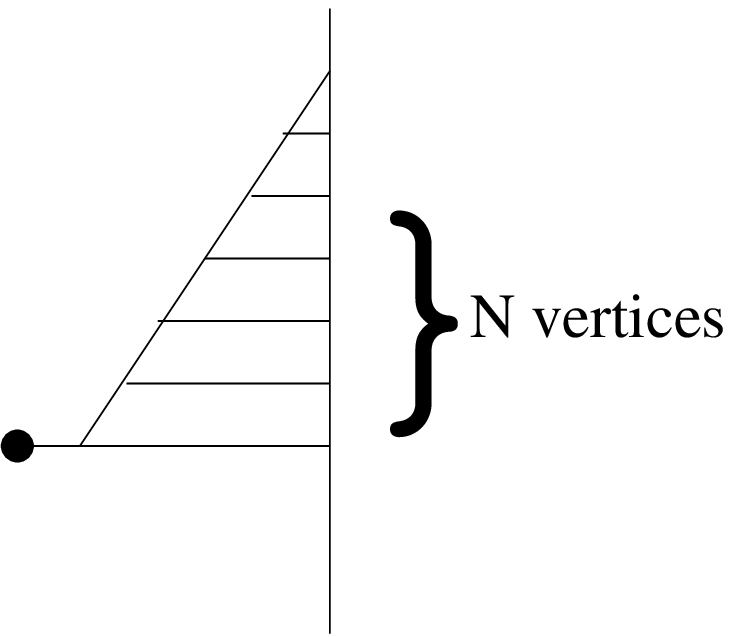}}\hfil\parbox{7cm}{
\epsfxsize 7cm\epsfbox{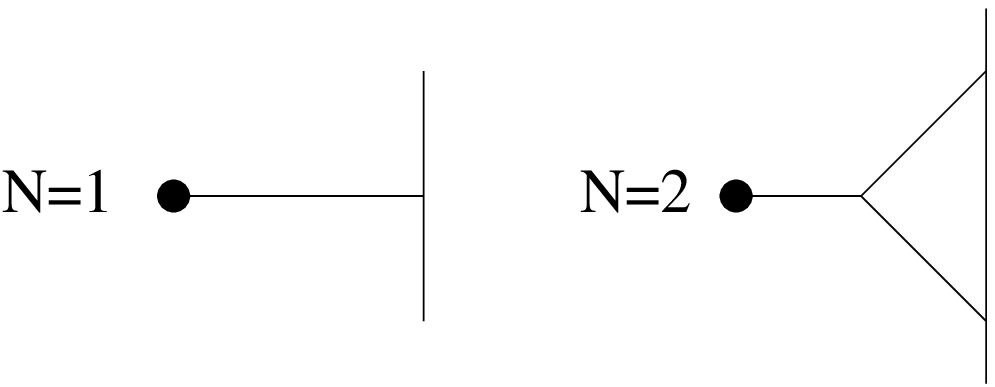}}
\caption{The ladder graph and the special cases for $N=1,2$.}
\label{ladder}
\end{figure}
For a given $N \geq 3$ there are precisely two of these, the one shown
in fig.\ref{ladder} and its mirror image. However, 
 there is only one graph for each of $N=1$ and $N=2$
 (fig.\ref{ladder}); we will define both of these to be ladders. All the
non-ladder graphs we will call ``trees''; so in fact there are no trees
for $N<4$.
Letting $L_N$ be the number of ladders with $N$ external vertices we
have
\bea L_1 = L_2 = 1 \nn \\
L_{N \geq 3} = 2 \nn \eea
so that
\beq L(z) = \frac{2 z}{1-z} - z - z^2 = z \left ( \frac{1+z^2}{1-z}
\right ) \label{lz} \eeq
and  the exponent $\gstr$ takes the value $2$ for a pure ladder
ensemble.
The number of trees $T'_N$ satisfies 
\bea T'_1 = T'_2 = T'_3 = 0 \nn \eea
\beq T'_N = q \left \{  \sum_{k=1}^{N-1} (T'_{N-k}+ L_{N-k})(T'_k + L_k) -
L_{N-1} - \delta_{N3} \right \},\quad N \geq 3. \label{tn} \eeq
The factor $q$ enables us to assign a different relative weight in the
ensemble to trees and ladders. A typical tree with $N$ external vertices 
gets a factor of
$q^{N-1}$; a tree which contains a ladder as a sub-graph has a smaller
power of $q$ associated with it. For the generating function we find 
\beq T'(z) = q \left \{ (T'(z) + L(z))^2 -z L(z) -z^3 \right \} \label{tz}
\eeq
which is easily solved to yield the generating function for the modified
ensemble
\beq {G}(z) = T'(z) + L(z) = \frac{1}{2q} \left ( 1- \sqrt{1-4 q \left (
(1-q z) L - q z^3 \right )} \right ). \label{gz} \eeq
For $q=1$ this just becomes the usual tree generating function
(\ref{tz}) with $\gstr=\frac{1}{2}$ whereas at $q=0$ it is equal to
$L(z)$. However for any positive non-zero value of $q$ we find that
$\gstr=\frac{1}{2}$. This is easily seen by considering the behaviour of
the argument of the square root as $z$ is increased from zero; as $z$
increases $L(z)$ increases but before it diverges the argument of the
square root must vanish (because it goes to $-\infty$ if $L(z)$ goes to
$\infty$). Thus only at $q=0$ exactly do we manage to ``freeze out'' the
general trees and get a system which contains only ladders. This
behaviour is very reminiscent of the $R^2$ model discussed in the
Introduction.

Of course this phase structure is consistent with the general arguments
we gave above. This is a system with one coupling constant, $q$, and for
generic values it has $\gstr=\frac{1}{2}$; at one particular coupling,
 $q=0$, it has $\gstr=2$, which
requires one constraint.

\section{Matter coupled to binary trees}

\begin{figure}[h]
\begin{center}\parbox{10cm}{
\epsfxsize 10 cm
\epsfbox{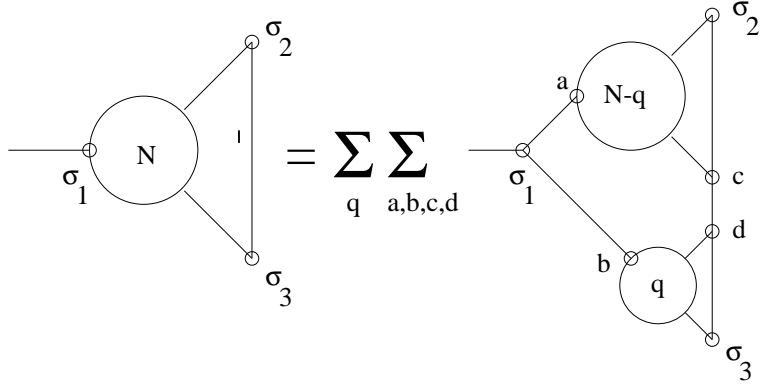}}\end{center}
\caption{Coupling of spins systems to a binary tree. A circle
containing $N$ represents graphs made of trees with $N$ external vertices
excluding the root. }
\label{iter2}
\end{figure}
We can extend the model by coupling matter to the trees.
Placing an
Ising spin $\sigma_{i} = \pm 1$ at each of the vertices we obtain the 
recurrence relation
shown in fig.\ref{iter2}.
The interaction  for two neighboring spins is given by $e^{\beta
\sigma_{i} \sigma_{j}}$. Using the identity $e^{\beta \sigma_{i} 
\sigma_{j}} = \cosh \beta [1 + t\sigma_{i}\sigma_{j}]$
where $t \equiv \tanh \beta$, we can take the link factor
to be $1 + t \sigma_{i} \sigma_{j}$. Then
\beq Z_{N} = \frac{1}{2^{4}} \sum_{q=1}^{N-1} \sum_{abcd} (1+t \sigma_{1}
a) (1+t \sigma_{1} c) (1+ t b d) Z_{N-q} (a, \sigma_{2}, b) Z_{q} (c, d,
\sigma_{3}). \label{z2} \eeq
In the absence of a magnetic field the dependence of $Z_N$ on the external
configuration $\sigma = (\sigma_{1}, \sigma_{2}, \sigma_{3})$ must take
the form
\beq Z_{N} (\sigma) = A_{N} + \sigma_{1} \sigma_{2} B_{N} + \sigma_{1}
\sigma_{3} B_{N} + \sigma_{2} \sigma_{3} C_{N}. \label{ansatz2} \eeq
Inserting (\ref{ansatz2}) into the partition function
(\ref{z2})  and equating coefficients depending on the same combination of
spins, we find a system of coupled equations for the
$A_{N}$,
$B_{N}$, $C_{N}$ in terms of $A_{N-1}$, $A_{N-2}$, etc. Defining the
grand-canonical partition functions 
\beq A (z,t) = \sum_{N=1}^{\infty} z^{N} A_{N} (t) \eeq
and similarly for $B (z,t)$, $C(z,t)$, we obtain the system of equations: 
\bea A &=& z + A^{2} + t^{3} B^{2} \label{z2in} \\
     B &=& z + t A B + t^2 B C \label{z2mid} \\
     C &=& z + t C^2 + t^2 B^2.\label{z2fin} \eea
Singularities in any of the functions $A$, $B$ or $C$ (or in any of their
derivatives) will signal a critical point in the full partition function
(\ref{z2}).

The simplest situation is when the matter is decoupled, $t=0$. Then $B = C
= z$ and $A(z)$ is just the tree generating function $T(z)$.
The system can also be solved exactly for $t = 1$. Then the equations
(\ref{z2in})-(\ref{z2fin}) imply that $A-C=A^2-C^2$ which only has one
sensible solution, $A=C$ (the other, $A=1-C$ does not exhibit the correct
behavior as $z \rightarrow 0$), which leads to $A=B$. Solving the
remaining quadratic we find
\beq A = B = C = \frac{1-\sqrt{1-8z}}{4}. \eeq
Hence, $z_c=\frac{1}{8}$ and $\gstr=\frac{1}{2}$. It is straightforward to
show that $\gstr=\frac{1}{2}$ for all $0 \leq t \leq 1$. Eliminating $A$
and $C$ from (\ref{z2in})-(\ref{z2fin}) gives
\beq B = \frac{z}{1-t+\frac{t}{2} \left [ \sqrt{1-4 z - 4 t^{3} B^{2}} 
+\sqrt {1 - 4 z t - 4 t^{3} B^{2}} \right ]}. \label{B} \eeq
Clearly $B$ itself cannot diverge as $z \uparrow z_{cr}$; this is
because the arguments of the square roots go negative when $B$ is still
finite. Hence $\gstr < 1$.
Differentiating with respect to $z$ we get
\bea  F \frac{\pa B}{\pa z} =
  \frac{B}{z} +\frac{B^{2} t}{z} \left \{ \frac{1}{\sqrt{1-4 z - 
4 t^{3} B^{2}}} + \frac{t}{\sqrt{1-4 z t - 4 t^{3} B^{2}}} \right \} \label{hh}
\eea
 where
\bea F =  1 - \frac{2 t^{4} B^{3}}{z} \left ( 
\frac{1}{\sqrt{1-4 z - 4 t^{3} B^{2}}} + \frac{1}{\sqrt{1-4 z t - 4 
t^{3} B^{2}}} \right )  \eea
The right hand side of (\ref{hh}) is clearly positive; the arguments
of the square roots cannot vanish as $z \uparrow z_{cr}$ because the
coefficient of $\frac{\pa B}{\pa z}$ would vanish while they were still
finite. So we are interested in the point where $F$
vanishes. $B$ is an increasing function of $z$ and grows faster than $z$
(by (\ref{B})). Hence $F$ is a monotonically decreasing function of $z$
at fixed $t$ and there is a value of $z$ for which $F=0$ and $\frac{\pa
B}{\pa z}$ diverges. Furthermore $\frac{\pa F}{\pa B}$ is a sum of
negative definite terms; it follows that the generic value
$\gstr=\frac{1}{2}$ applies everywhere and the cancelation necessary to
get any other value of $\gstr$ cannot occur.

The critical structure 
of the theory is not changed by the addition of Ising spins. This result
is expected because the Ising spins never have a diverging correlation
length in less than two dimensions and so cannot affect the global
properties of the geometry.

We now consider a generalised Ising model in which the  spins take the
values
$1$, $ e^{\pm \frac{2 \pi i}{3}}$ \cite{mirza}. 
By allowing the partition function to include complex weights, 
we obtain a richer phase structure than that of the 2-state case. 
The partition function is given by
\beq Z(\alpha, \beta) = \sum_{{s}} e^{\alpha \sum_{(ij)} 
(s_i^{\dagger} s_j - s_i s_j^{\dagger}) +\beta \sum_{(ij)} 
(s_i^{\dagger} s_j + s_i s_j^{\dagger})} \label{zcomplex} \eeq 
(note that in doing this it is necessary to assign an orientation to every
link). The Boltzman factor for a link can be written in the form
\beq \tilde L = \lambda (1+t(s_i^{\dagger} s_j + s_i s_j^{\dagger}))
(1+ \gamma (s_i^{\dagger} s_j - s_i s_j^{\dagger})) \eeq
where, 
\bea \lambda &=& \left [(1+\gamma^2)(1-t^2)(1+2t)\right]\nn \\
\gamma &=& \frac{\tan(\sqrt{3} \alpha)}{3} \nn \\
 t &=& \frac{e^{3 \beta} - \cos(\sqrt{3} \alpha)}{2 \cos(\sqrt{3} \alpha) 
+ e^{3 \beta}} . \eea
We see from these expressions for that our model will be well defined for
$\lambda$ in the interval $[0, +\infty]$; however $t$ must be restricted to
$[0,1[$ for $\lambda$ to be analytic. 
We can thus drop $\lambda$ and consider
\beq L = (1 + \mu S_{i} S_{j}^{\dagger} + \nu S_{i}^{\dagger} S_{j}) \eeq
where $\mu \equiv t + \gamma (1-t)$ and $\nu \equiv t-\gamma (1-t)$ so
that the partition function (\ref{zcomplex}) becomes
\beq \CZ = \sum_{trees} Z_{N} = \sum_{trees} \sum_{\{S_{i}\}} \prod_{links} 
(1 + \mu S_{i} S_{j}^{\dagger} + \nu S_{i}^{\dagger} S_{j}). \eeq
We can proceed as before to obtain a 
system of coupled equations which completely describe $Z$. The most general 
form of $Z_N$, taking into account its global ${\bf Z}_{3}$ symmetry, is
\begin{eqnarray} Z_{N} (S_{1}, S_{2}, S_{3}) = A_{N} + S_{1} S_{2}^{\dagger} 
B_{N} + S_{1} S_{3}^{\dagger} \tilde B_{N} + S_{1}^{\dagger} S_{2} D_{N} + 
S_{1}^{\dagger} S_{3} \tilde D_{N} +\nn \\ S_{2} S_{3}^{\dagger} C_{N} + S_{3} 
S_{2}^{\dagger} E_{N} + S_{1} S_{2} S_{3} F_{N} + S_{1}^{\dagger} 
S_{2}^{\dagger} S_{3}^{\dagger} G_{N}. \label{A} \end{eqnarray}
By introducing this form into the partition function and then computing
the sums leading to the grand canonical partition function, we arrive at
the following set of equations: 
\bea A &=& z + A^2 + \mu \nu (\mu + \nu) BD \label{z3in} \nn\\
 B &=& z + \mu \tilde B A + \mu \nu B C + \nu^{2} \mu DG\nn \\
 \tilde B &=& z + \mu \tilde B A + \mu^{2} B C + \nu^{3} D G\nn  \\
 C &=& z + \mu C E + \mu \nu \tilde B \tilde D + \nu^{2} \mu F G\nn  \\
 D &=& z + \nu A \tilde D + \mu \nu D E + \mu^{2} \nu B F\nn  \\
 \tilde D &=& z + \nu A \tilde D + \nu^{2} D E + \mu^{3} B F\nn  \\
 E &=& z + \nu C E + \mu \nu \tilde B \tilde D + \mu^{2} \nu F G\nn  \\
 F &=& z + \mu (\mu +\nu) E F +\nu^{2} \tilde D^{2}\nn  \\
 G &=& z + \nu (\mu +\nu) C G + \mu^{2} \tilde B^{2}. \label{z3fin} \eea
For some curves in the $(t$,$\gamma)$ plane this set of equations can be
solved exactly. We will find that, taken together with the general discussion 
of section 1, this is sufficient to enable us to
elucidate the entire phase diagram.

At $t=\gamma=0$, $A$ is given as usual by the
generating function of the Catalan numbers and all other quantities equal
$z$ so $\gstr=\frac{1}{2}$. At $\gamma=0$ (this is the 3-state Potts
model) we have that $\mu=\nu=t$ and the equations can be reduced by
elementary manipulations to the set
\bea A &=& z + A^2 + 2 t^3 B^2 \nn \\
     B &=& z + t B A + t^2 B C + t^3 B F \nn \\
     C &=& z + t C^2 + t^2 B^2 + t^3 F^2 \label{system1} \\
     F &=& z + 2 t^2 C F + t^2 B^2 \nn \\
     \tilde B &=& B,\:  \tilde D = D,\: E = C,\:  G = F, \nn \eea
which, like the Ising model (\ref{z2in})-(\ref{z2fin}), can be shown to have 
$\gstr=\frac{1}{2}$ for all $z$. At $t=1$,
$\mu=\nu=1$ independently of $\gamma$ and the system is again described by 
(\ref{system1}) having $\gstr=\frac{1}{2}$ for all $\gamma$; in this case
the equations have elementary solutions
\bea A &=& \frac{1}{2} \left ( 1- \sqrt{ \frac{5}{9}-\frac{4}{3} z + 
\frac{4}{9} \sqrt{1 - 12 z}} \right ) \nn \\
B &=& \sqrt{\frac{1 - 6 z - \sqrt{1 - 12 z}}{18}} \label{sols1} \\
C &=& \frac{1}{4} \left ( 1- \sqrt{ \frac{5}{9} - \frac{16}{3} z +
\frac{4}{9} \sqrt{1 - 12 z}} \right ) \nn \\
F &=& C \nn \eea
and all the functions are non-analytic at $z_{cr} = \frac{1}{12}$.

Along the line $t= \frac{\gamma}{1+ \gamma}$, $\nu=0$ and the system of
equations is greatly simplified and can be solved to give
\bea A &=& 
\frac{1}{2} \left ( 1 -  \sqrt{1- 4 z} \right ) \nn \\
\tilde B &=& \frac{z (1 - \mu z) + \mu^2 z^2}{1-\mu z - \mu (1 - \mu z +
\mu^2 z)\frac{1}{2} \left ( 1 -  \sqrt{1- 4 z} \right )} \nn \\
B &=& z + \mu \tilde B A \nn \\
C &=& \frac{z}{1- \mu z} \nn \\
D &=& E = z \nn \\
\tilde D &=& z + \mu^3 B F \nn \\
F &=& \frac{z}{1-\mu^2 z} \nn \\
G &=& z + \mu^2 \tilde B^2.\label{sols} \eea 
At small $\mu$ the singularity is at $z=\frac{1}{4}$ and $A$, $B$, $\tilde
B$, $G$ are non-analytic with $\gstr=\frac{1}{2}$ while $C$, $D$, $E$, $F$
are analytic. So for small $\mu$ the $\nu=0$ line lies within the 
$\gstr=\frac{1}{2}$ region and is not a phase transition. 
However as $\mu$ increases we reach a point where the
denominator of $\tilde B$ vanishes for $z < \frac{1}{4}$; this occurs at
$\mu \geq \mu_0$ where
\beq 1 - \frac{3}{4} \mu_0 + \frac{\mu_0^2}{8} - \frac{\mu_0^3}{8} = 0
\label{muzero} \eeq
which corresponds to $\mu_0 \simeq 1.2633$, i.e. $t \simeq 0.6316$.
So at $\mu = \mu_0$ the singularity is at $z= \frac{1}{4}$ but now
\beq \tilde B \sim \frac{1}{\sqrt{1-4 z}} \eeq
and at this point $\gstr=\frac{3}{2}$. For $\mu > \mu_0$ we have that
\beq \tilde B \sim \frac{1}{z_{cr}-z} \eeq
with $z_{cr} < \frac{1}{4}$ so that A becomes analytic and $\gstr=2$.
From our general considerations we expect that $\gstr=\frac{3}{2}$ 
occurs at a point and $\gstr=2$ along a line in this two-dimensional
coupling constant space. Hence for $\mu > \mu_0$ the line $\nu>0$ separates
two phases.

\begin{figure}[h]
\parbox{14cm}{\epsfxsize 14 cm
\epsfbox{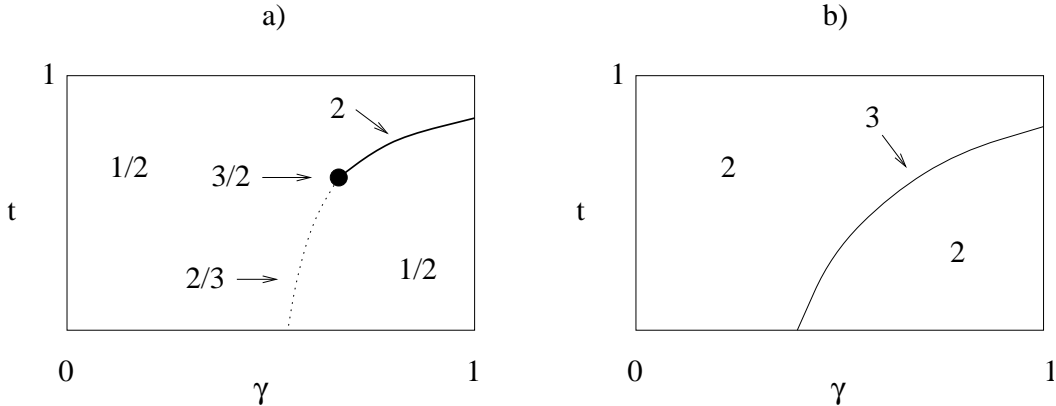}}
\caption{The phase diagram for a) the complex Ising coupled to binary trees,
and b) the complex Ising coupled to the ladder.
Numbers indicate values of $\gstr$.}
\label{phased1}
\end{figure}

It is straightforward  to show from (\ref{z3fin})that for $\mu\ne 0$,
$\nu\ne 0$, all the potentials are finite (this is done by assuming
that a potential diverges at the critical point and obtaining a contradiction).
Thus the two phases which are separated by the $\gstr=2$ line must both have
$\gstr=\frac{1}{2}$, this being
the only value which can exist on an area of the phase diagram of a system
with two coupling constants.  The nature of the phases is quite different.
For $t>t_0$ and $\nu>0$ the thermodynamic limit exists in the usual sense, e.g.
\beq \lim_{N\to\infty} \frac{1}{N}\log A_N =const.\label{tdl}\eeq
On the other hand when $t>t_0$ and $\nu<0$ the potentials fluctuate in sign 
with increasing $N$; $z_{cr}$ has become complex. Of course this fluctuation is a remnant of the complex
(and sometimes negative) weights in the partition function.  For $t<t_0$
the line $\nu=0$ falls in the phase with the conventional thermodynamic limit
as can be seen from the solutions (\ref{sols}). The line separating the
two phases must be in the $\nu<0$ region and, because all the potentials
are finite, must therefore have $\gstr=\frac{2}{3}$ thus completing the
phase diagram shown in fig.\ref{phased1}a.

Now consider the behaviour of the model when we weight ladder
configurations differently from the rest of the ensemble as in section 3. 
At $q=0$ only
ladder configurations survive; in the thermodynamic limit they are
effectively one-dimensional objects, and it is straightforward to compute
the transfer matrix to obtain the phase diagram depicted in fig.4b.
To elucidate the general structure of the phase diagram for the three coupling
constants $(\nu \mbox{,} \mu \mbox{,} q)$, one would in principle repeat
the procedure that lead to equations (\ref{lz})-(\ref{gz}).
However it is clear that the
general structure of the equations  will be a system of
coupled quadratic equations of the form discussed in section 2. Hence we
can extend the analysis performed there to the present case.

It is clear that the plane $q=0$ is exceptional; in this plane the coupled
equations are in fact linear which is why $\gstr$ takes integer values.
Viewed as part of a   three coupling
constant $(\nu \mbox{,} \mu \mbox{,} q)$ space, $\gstr=2$ can only
exist on a plane  and $\gstr=3$ only
on a line. On the other hand of the values for $\gstr$ which occur at $q=1$,
 $\gstr=\frac{1}{2}$ can exist on a volume,
$\gstr=\frac{3}{2}$ on a line, and $\gstr=\frac{2}{3}$ on a plane.
If we move from $q=0$ to any positive $q$, we cannot have
$\gstr=2$ on a volume and hence the two regions which have that value at
$q=0$ disappear for any positive $q$; the same happens to the line $\gstr=3$.
The general form of the phase diagram for $q=1$ is mantained for any
positive non-zero value of $q$.

This situation mirrors the case of no matter coupling closely: there is an 
exceptional plane $q=0$ where, by virtue of the ``freezing'' of the
configurations, the values $\gstr$ takes are different from the ones at
positive non-zero $q$. As soon as one moves away from that plane, one is
in a region where all configurations are important for the
thermodynamics of the ensemble, and the results are qualitatively
similar to those obtained at $q=1$ where there is no damping of
non-ladder configurations.

\section{Conclusions}

In this paper we investigated  models of matter coupled to
binary trees. Coupling the Ising model
to such a system does not alter its critical properties, as might be
 expected
 but by
 introducing a complex-action ${\bf Z}_3$  model we showed that new behaviour 
can arise, giving values of $\gstr$
different from $\frac{1}{2}$. Interestingly, for a specific set of
coupling constants $\nu=0$, $t>t_0$, we find $\gstr=2$ which is the 
behaviour expected when  one single type of linear configuration
dominates the full partition function; this is a sort of 
dimensional collapse driven by interaction with matter fields,
 although of a different sort from that studied in 
\cite{me}. On the other hand it is also possible to formulate these models 
to study explicitly the transition between a fixed configuration, the ladder, 
and the full fluctuating ensemble. We found that only when fluctuations
are totally suppressed do the critical exponents change  relative to
the gravity case. 

\vspace{1 truecm}
\noindent {\bf Aknowledgments: }J.D.C. acknowledges  a grant from 
{\sc Praxis XXI}.

\end{document}